\documentclass[aps,prc,twocolumn,preprintnumbers,groupedaddress,superscriptaddress]{revtex4}
\usepackage{graphicx}
\usepackage{color}

\newcommand{\pT}{$p_T$ }
\newcommand{\mT} {$m_T$ }

\newcommand{\sNN}{$\sqrt{s_{_\mathrm{NN}}}$ }
\newcommand{\s}{$\sqrt{s}$ }
\newcommand{\pp}{$p$+$p$ }

\begin{document}

\preprint{Intend to PRC(R), version 3}

\title{Spectra and radial flow at RHIC with Tsallis statistics in a Blast-Wave description}


\author{Zebo Tang}
%\email{zbtang@mail.ustc.edu.cn}
%\thanks{}
\affiliation{University of Science \& Technology of China, Hefei
230026, China}
\author{Yichun Xu}
\affiliation{University of Science \& Technology of China, Hefei
230026, China}
\author{Lijuan Ruan}
\affiliation{Brookhaven National Laboratory, Upton, New York
11973, USA}
\author{Gene van Buren}
\affiliation{Brookhaven National Laboratory, Upton, New York
11973, USA}
\author{Fuqiang Wang}
\affiliation{Purdue University, West Lafayette, Indiana 47907,
USA}
\author{Zhangbu Xu}
\email{xzb@bnl.gov} \affiliation{Brookhaven National Laboratory,
Upton, New York 11973, USA}


\date{\today}

\begin{abstract}
We have implemented the Tsallis statistics in a Blast-Wave model and
applied it to mid-rapidity transverse-momentum spectra of identified particles
measured at RHIC. This new Tsallis Blast-Wave function
fits the RHIC data very well for $p_T<$3 GeV/$c$. We observed that the
collective flow velocity starts from zero in p+p and peripheral Au+Au
collisions growing to 0.470 $\pm$ 0.009($c$) in central Au+Au
collisions. The $(q-1)$ parameter, which characterizes the degree of
non-equilibrium in a system, changes from $0.100\pm0.003$ in p+p to
$0.015\pm0.005$ in central Au+Au collisions, indicating an evolution
from a highly non-equilibrated system in p+p collisions toward an
almost thermalized system in central Au+Au collisions. The temperature
and collective velocity are well described by a quadratic dependence
on $(q-1)$. Two sets of parameters in our Tsallis Blast-Wave model are
required to describe the meson and baryon groups separately in p+p
collisions while one set of parameters appears to fit all spectra in
central Au+Au collisions.
\end{abstract}

%\keywords{}

\maketitle


\section{Introduction}

Identified particle spectra in transverse momenta provide pillars in
the discoveries in relativistic heavy ion
collisions~\cite{starWhitePaper,phenixWhitePaper,Gyulassy:2004zy,muller:2004kk}.
In Au+Au collisions at RHIC, identified particle yields per unity of
rapidity integrated over the transverse momentum range have provided
information on chemical potential and temperature at the chemical
freeze-out in a statistical
analysis~\cite{starWhitePaper,starPiKPLowPtPRL}. The transverse
momentum (\pT) distributions of particles with different masses can be
described in a Boltzmann-Gibbs Blast-Wave (BGBW) model with a
compact set of parameters of temperature ($T$), flow velocity ($\beta$ )
and flow profile ($\rho$)~\cite{starWhitePaper,blastWave}. With the
BGBW model applied to the data in a limited \pT range, a large radial
flow $\beta\simeq$ 0.6 is obtained in central Au+Au collisions, while
much smaller flow is obtained in peripheral Au+Au
collisioins~\cite{starPiKPLowPtPRL,Retiere_Lisa_PRC70}. The
description of heavy ion collisions using these freeze-out conditions
have been used as evidence of collective motion in relativistic heavy
ion
collisions~\cite{starWhitePaper,phenixWhitePaper,voloshin:2008dg,Adams:2003fy,Adler:2003cb}
(which has further allowed extraction of drag and diffusion
coefficient of heavy quarks in the expanding
bulk~\cite{Teaney:2006ec,rapp:2008qc,starCharm:2008hja,Zhao:2008vu}),
provided necessary interplay with elliptic flow effect for the
observed mass ordering in $v_2$~\cite{voloshin:2008dg}, and as a
necessary condition for explaining the ridge phenomenon in many
models~\cite{voloshin:2003ud,Armesto_jetquenching05,Dumitru:2008wn,Gavin:2008ev,Shuryak:2007fu}.

However, such BGBW descriptions have limitations. Modeling nuclear
collisions in ideal hydrodynamics is limited to low \pT ($\lesssim$ 1
GeV/$c$) because it is generally believed that the equilibrium
description fails at high $p_T$, where particle production may be
dominated by non-equilibrium or hard processes and exhibits a
characteristic power-law tail~\cite{Wilk:1999dr}. The blast-wave model
has a strong assumption on local thermal equilibrium so that a
Boltzmann distribution can be applied~\cite{blastWave}. This results
in an arbitrary choice of \pT range of the spectra where the function
is able to fit the data and requires low and high \pT
cuts~\cite{starPiKPLowPtPRL,Retiere_Lisa_PRC70}. The BGBW model also
lacks non-extensive quantities to describe the evolution from p+p to
central A+A collisions. For example, one would expect that the energy
deposited at mid-rapidity for particle production fluctuates
significantly from event to event in p+p collisions while the current
BGBW treats all p+p events the same with a heat bath at a fixed
temperature. The resulting finite flow velocity $\beta \simeq$ 0.2 in
p+p collisions from the same fitting procedure obscures the
interpretation that collective flow is large and unique in A+A
collisions~\cite{starPiKPLowPtPRL}. In Au+Au collisions, the
fluctuations at initial impact due to Color-Glass Condensate (CGC)
formation or individual nucleon-nucleon collision may not be
completely washed out by subsequent interactions at either the QGP
phase or hadronic
phase~\cite{Drescher:2000ec,Mishra:2007tw,Broniowski:2007nz}. All
these effects leave footprints in the spectra at low and intermediate
\pT ($p_T \lesssim$ a few GeV/$c$). For example, the volution of
$\phi$ and $K^*$ spectra from an \mT power-law (Levy) function in p+p
and peripheral Au+Au collisions to an \mT exponential (Boltzmann)
function in central Au+Au collisions can be clearly
observed~\cite{starKstar130,starKstar,Adams:2004ux,starPhiAuAu200},
where $m_T=\sqrt{m_{0}^{2}+p_T^{2}}$ is the transverse mass of 
particle with mass $m_0$ at a given \pT.

With its development and success of Tsallis statistics in dealing with
complex systems in condensed matter, many authors have utilized
Tsallis statistics to understand the particle production in
high-energy and nuclear
physics~\cite{De:2007zza,Wilk:2008ue,Alberico:1999nh,Osada:2008sw,Biro:2003vz}.
Although the implications and understanding of the consequences of
such an application are still under investigation, the function is
relatively easy to understand. The usual Boltzmann distribution in an
\mT exponential form is re-written as an \mT power-law function:
\begin{equation}
\frac{d^2N}{2\pi m_T dm_T
dy}\propto(1+\frac{q-1}{T}m_T)^{-1/(q-1)}\label{levy function}
\end{equation}
where the left-hand side is the invariant differential particle yield
and $q$ is a parameter characterizing the degree of
non-equilibrium. The distribution can be derived from the usual
procedure in statistical mechanics, starting from a non-equilibrium
$q$-entropy~\cite{q-entropy}. If particle production is not
distributed at a fixed temperature but rather as a system where $T$
varies with total energy (E) as $T \propto (q-1)E$, this will result
in a negative binomial distribution (NBD) for particle number and
temperature fluctuations.  Their relations to $q$ are the skewness in
NBD $\kappa=1/(q-1)$ and the fluctuation of temperature
$\sigma^2(1/T)=(q-1) \langle 1/T \rangle ^2$~\cite{beck}. Regardless
of the physical interpretation, Eq.~\ref{levy function} does provide a
necessary power-law behavior at high \pT and an exponential behavior
at low \pT. This is exactly what has been observed in p+p and
peripheral Au+Au collisions at RHIC for $K^*$ and $\phi$, whose
measurable \pT range can reach low and high
\pT~\cite{starKstar130,starKstar,Adams:2004ux,starPhiAuAu200}. When
$q\rightarrow1$, Eq.~\ref{levy function} becomes the familiar
Boltzmann distribution again.

In addition to the features of the spectral evolution necessary to
describe the \pT spectra observed at RHIC, the physical interpretation
of the results can provide quantitative insight into the
non-equilibrium processes in relativistic heavy ion collisions and
whether the system has thermalized in central Au+Au collisions. We
expect that individual nucleon-nucleon collisions inside a
nucleus-nucleus collision are in non-equilibrium at the initial impact
and/or have a large energy fluctuation or a large $(q-1)$ value, which
produce a large power-law tail in the \mT spectra. Alternatively, in a
CGC scenario, the system as a whole produces a strong color field with
large
fluctuations~\cite{Dumitru:2008wn,Gavin:2008ev,Sorensen:2008bf}. This
can be treated as many hot spots in a nucleus-nucleus collision at the
initial impact. In a viscous hydrodynamic evolution, the hot spots are
smoothed (dissipated) into producing collective flow, creating more
particles and increasing temperature~\cite{hydroReview}. It has been
argued in the Tsallis statistics that the increase of temperature and
flow velocity during the evolution is connected to the decrease of
$(q-1$) by (shear and bulk $\xi$) viscosity in linear or quadratic
proportion~\cite{Wilk:1999dr,Wilk:2008ue}, such as
$T=T_0+a\xi(q-1)^2$. This can provide quantitative insight into the
bulk viscosity, which is predicted to peak at the phase transition and
is much larger than the shear
viscosity~\cite{Kharzeev:2007wb,Karsch:2007jc}.

In this paper, we present the procedure of implementing Tsallis
statistics in the Blast-Wave model (TBW) and use it to fit the
identified particle spectra at mid-rapidity at RHIC. The physics
implications are discussed and preparation for future work is also
presented. Good TBW fits can also provide a practical experimental
tool to extract particle yields ($dN/dy$) by extrapolating to
unmeasured kinematic ranges since most of the experimental
measurements only cover a limited \pT range for any given particle.

\section{Implement Tsallis Statistics into Blast-Wave model}

To take into account collective flow in both longitudinal and
transverse directions in relativistic heavy ion collisions, a simple
Tsallis distribution needs to be embedded in the framework of
hydrodynamic expansion~\cite{Wilk:2008ue}. We follow the recipe of the
Blast-Wave model provided by Schnedermann et al.~\cite{blastWave}. The
formula has been adopted by many authors to implement a fit to the
data in relativistic heavy ion collisions
~\cite{Retiere_Lisa_PRC70,starPiKPLowPtPRL,Adler:2003cb,fuqiangLongPID}. It
is relatively trivial to change sources of particle emission from a
Boltzmann distribution to a Tsallis distribution in the Blast-Wave
model.
\begin{widetext}
\begin{equation}
\frac{dN}{m_T dm_T} \propto m_T \int_{-Y}^{+Y}\cosh(y)dy
\int_{-\pi}^{+\pi}d\phi
\int_{0}^{R}rdr(1+\frac{q-1}{T}(m_T\cosh(y)\cosh(\rho)-p_T\sinh(\rho)\cos(\phi)))^{-1/(q-1)},
\end{equation}
\end{widetext}
where $\rho = \tanh^{-1}(\beta_s(\frac{r}{R})^n)$ is the flow profile
growing as $n$-th power from zero at the center of the collisions to
$\beta_s$ at the hard-spherical edge ($R$) along the transverse radial
direction ($r$), and $\beta=\beta_s/(1+1/(n+1))$ is the average flow
velocity. We have used $n=1$ in this study. However, the integrations
after the replacement are hypergeometric functions and the integrals
over rapidity ($y$) and azimuthal angle ($\phi$) cannot be decoupled
into two Bessel functions as in BGBW~\cite{blastWave}. The computing
program typically used to provide conventional Blast-Wave fits to RHIC
data~\cite{fuqiangLongPID} has been modified to numerically calculate
the integration in the above spectral function and fit the reslting
distributions to the data. There are a few assumptions which do not
change when a Tsallis distribution replaces a Boltzmann distribution:
\begin{enumerate}
    \item {Bjorken longitudinal expansion is assumed so that the
measured particle yield does not depend on rapidity due to the
integration over rapidity of the source~\cite{blastWave}. This is
approximately true at mid-rapidity at RHIC or
LHC~\cite{BRAHMSrapidity}. On the other hand, this assumption can be
lifted in future analyses with a more complicated integration provided
an emitting source along rapidity is known.}
    \item Isotropic emission in azimuth is assumed for each local
source. However, the distribution of the source can have an azimuthal
dependence in reality~\cite{starv2}. In the future, this can be
implemented as has been done in the conventional BGBW model to fit the
azimuthal dependence of HBT radius and the identified particle
elliptic flow~\cite{Retiere_Lisa_PRC70}.
    \item The emission source everywhere has the same density and
degree of non-equilibrium ($q$) at the time of kinetic
freeze-out. This may not be true since the high-$p_T$ particles (jet)
tend to have surface
emission~\cite{Loizides:2006cs,Zhang:2007ja}. This kind of corona
effect has been implemented in many other models and can be adopted in
this framework as well.
  \item Resonance decay contributions to the stable particle yields
  have been treated as part of the source emission. The detailed decay
  kinematics and its effect on the spectra have been studied in
  Ref.~\cite{blastWave,fuqiangLongPID}. Incorporation of resonance
  effects on spectra can be made in future improvements.

\end{enumerate}

The goal of this paper is to provide a first implementation of the TBW
model to exercise a trial case with the RHIC data. Future work can
change the above assumptions to resemble more realistic conditions by
comparing the assumptions to data and hydrodynamic calculations.

\begin{figure}[ht]
\includegraphics[width=0.49\textwidth]{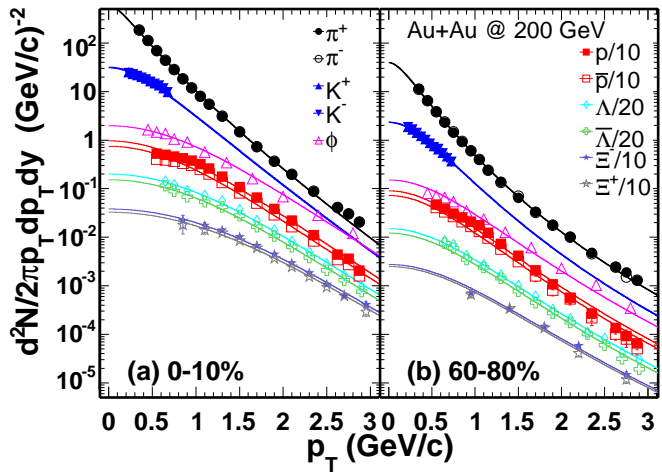}
\caption{(Color Online) Identified particle transverse momentum
spectra in Au+Au collisions at \sNN = 200 GeV in0-10\% central (a) and
in peripheral 60-80\% collisions (b).  The symbols represent experiment
data points. The solid curves represent the TBW fit.}
\label{fig:spectraFitAuAu}
\end{figure}

\section{Results of fit to RHIC data}

The STAR Collaboration has published a series of particle spectra at
mid-rapidity. The most complete set is for p+p and Au+Au collisions at
\sNN= 200 GeV. The identified particle spectra include $\pi^{\pm}$,
$K^{\pm}$, $K_S$, $K^*$, $p$, $\phi$, $\Lambda$, $\Xi$, $\bar{p}$,
$\bar{\Lambda}$, and
$\bar{\Xi}$~\cite{starWhitePaper,starKstar,starPhiAuAu200,ppSplit,
Abelev:2006jr,Abelev:2007ra,Adams:2003qm,Adams:2004ux,
Adams:2006ke,Adams:2006nd}. The invariant differential yields were
measured as $\frac{d^2N}{2\pi p_Tdp_Tdy}$. In the new TBW model, three
parameters are common for all particles: temperature $T$,
non-equilibrium parameter $q$, and maximum flow velocity $\beta_s =
\beta(1+n/2)$ where $n=1$ and average flow velocity $\beta$ is bounded
to the range $[0,0.7]$~\cite{fuqiangLongPID} to aid in fit convergence
and avoid non-physical results. An additional parameter provides the
overall normalization of $dN/dy$ for each species. We choose the
Minuit in Root~\cite{Minuit} to perform a least-$\chi^2$ fit used in
ref.~\cite{starWhitePaper}.  Figure~\ref{fig:spectraFitAuAu} shows the
\pT spectrum data together with our fit results in two selected
centrality bins (0-10\% and 60-80\%) in Au+Au collisions. The fit
parameters and $\chi^2$/DoF are tabulated in Tab.~\ref{table}. As
stated earlier in our model's third assumption, surface emission could
become important at high $p_T$; we limit our fits to $p_T<3$ GeV/$c$
to avoid this region, which still extends the fit range well beyond
previous BGBW fits. The curves from our model generally describe the
data very well, especially in central Au+Au collisions. For peripheral
Au+Au collisions, the meson spectra are well described by the model
while the baryons are in general over-predicted at higher \pT. On the
other hand, the $\chi^2$/DoF show good fits in all cases. The main
results are:
\begin{enumerate}

        \item
        ($q-1$), a measure of the degree of non-equilibrium, decreases by a
        factor of 5 from 0.086 to 0.018. This means the power in the
        \mT power-law increases from about 12 to 56, attaining an almost Boltzman
        distribution.

        \item
        $T$, the average temperature of the local source, shows a
        small increase from 114 MeV to 122 MeV. This trend is in
        contrast to the conventional BGBW result, where a decrease of
        temperature was observed~\cite{fuqiangLongPID}.

        \item
        $\beta$, the average flow velocity, increases from 0 in peripheral to
        0.47$c$ in central Au+Au collisions. That the minimum
        $\chi^2$/DoF is found at the lower bound of 0 for
        peripheral collisions indicates that either the model is
        incomplete (some approximations may not be sufficiently true),
        or that no flow has developed in peripheral collisions within
        the context of the model. This also coincides with a large
        q-1, indicating a very non-equilibrated system if the
        description as a unified system applies.

\end{enumerate}

\begin{table}%[H] add [H] placement to break table across pages
\caption{Values of parameters from TBW fit to identified particle
transverse spectra in Au+Au collisions of different centralities and
in \pp collisions at RHIC. Quoted errors are quadratical sum of
statistical and uncorrelated systematic errors. The limits of $\beta$
is set to $[0,0.7]$. \label{table}}
\begin{ruledtabular}
\begin{tabular}{ccccc}
centrality & $\beta$ &  T  & $q-1$ & $\chi^2$/nDoF\\
\hline
0-10\%   & 0.470$\pm$0.009 & 0.122$\pm$0.002 & 0.018$\pm$0.005 & 130/125\\
10-20\%  & 0.475$\pm$0.008 & 0.122$\pm$0.002 & 0.015$\pm$0.005 &  119/127\\
20-40\%  & 0.441$\pm$0.009 & 0.124$\pm$0.002 & 0.024$\pm$0.004 &  159/127\\
40-60\%  & 0.282$\pm$0.017 & 0.119$\pm$0.002 & 0.066$\pm$0.003 &  165/135\\
60-80\%  & $0^{+0.05}_{-0}$ & 0.114$\pm$0.003 & 0.086$\pm$0.002 &  138/123\\
Meson pp  &    0           & 0.089$\pm$0.004 & 0.100$\pm$0.003 &  53/66\\
Baryon pp &    0           & 0.097$\pm$0.010 & 0.073$\pm$0.005 & 55/73\\
% Lines of table here ending with \\
\end{tabular}
\end{ruledtabular}
\end{table}

\begin{figure}[ht]
\includegraphics[width=0.41\textwidth]{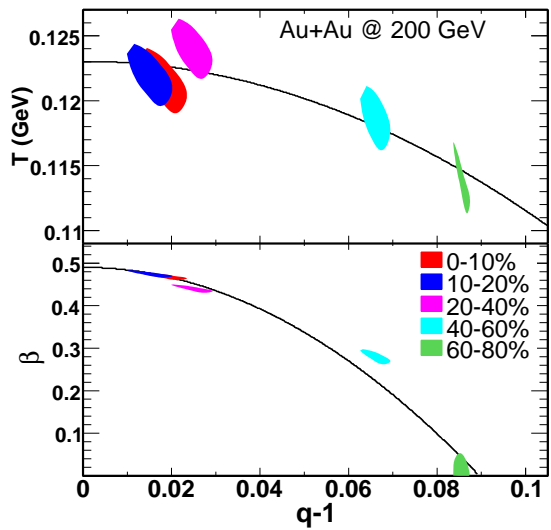}
\caption{(Color Online) The fit parameters $T$ and $\beta$ as a
function of $(q-1)$. Each block is one-$\sigma$ contour from the error
matrix of the TBW fit for a given centrality of Au+Au collisions.}
\label{fig:contour}
\end{figure}

Figure~\ref{fig:contour} shows the temperature and flow velocity
versus ($q-1$) for Au+Au collisions. Each shaded region represents a
one-$\sigma$ contour from the error matrix obtained from the TBW fit
for a given centrality. The dependence is clearly non-linear and has a
negative correlation. There is a jump of flow velocity from zero in
p+p and 60-80\% Au+Au centrality to 0.28 at 40-60\% Au+Au centrality,
coinciding with the transition behavior in several other
observables~\cite{Sorensen:2008bf}. We fit the distributions with a
quadratics and obtain $T=(0.123\pm0.0014)-(1.2\pm0.4)(q-1)^2$ and
$\beta=(0.49\pm0.01)-(61\pm5)(q-1)^2$, as shown in the figure.

\begin{figure}[ht]
\includegraphics[width=0.49\textwidth]{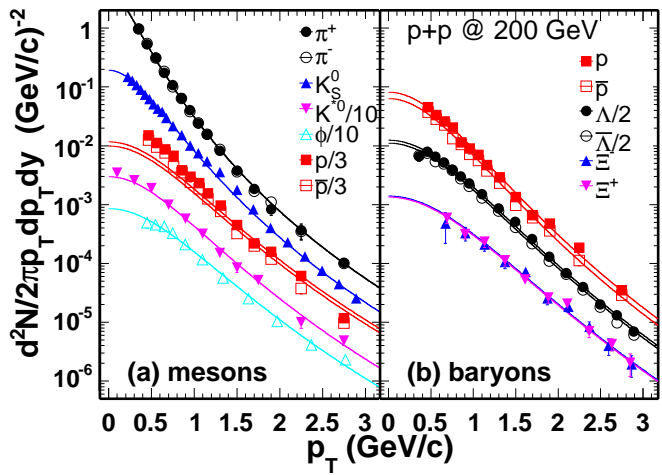}
\caption{(Color Online) Transverse momentum spectra of identified
mesons (a) and baryons (b) produced in \pp collisions at \s = 200
GeV. The symbols represent the experiment data. The solid curves
represent the TBW fit. The proton and anti-proton transverse momentum
spectra and predicted curves from meson fit parameters are also
plotted on panel (a) for comparison.} \label{fig:spectraFitpp}
\end{figure}

Since the TBW model can be used to describe systems at
non-equilibrium, it is natural to extend the fit to p+p
collisions. However, a very poor $\chi^2$/DoF was obtained if we
include all of the mesons and baryons in a common fit. Instead, two
separate groups of mesons and baryons show good fits.
Figure~\ref{fig:spectraFitpp} shows the results of the fits together
with the data points. In both cases, the flow velocity was set at the
lower limit of $\beta=0$ as was also independently verified if $\beta$
was set to be a free parameter. In Fig.~\ref{fig:spectraFitpp}.a, the
proton and anti-proton spectra are presented together with the
predicted curves from the TBW fit to the meson group. The need to
separate mesons and baryons can be seen from spectral shape of the
proton data, which matches more closely that of the kaons than the
$K^*$ (steeper at low \pT) despite being closer in mass to the
$K^*$. The TBW model restricts spectral shapes to vary monotonically
with mass for fixed ($T$, $q$, $\beta$), so there is no allowance for
this artifact. Weak-decay effects from $\Lambda$ are investigated
based on the results from the baryon-only fit and $\Lambda-$decay
kinematics and do not explain this large steepening. On the contrary,
a significantly smaller $q$ flattens the spectra at low \pT and
softens them at high \pT, which is necessary and sufficient to achieve
a good fit for the baryons ($p$, $\Lambda$, $\Xi$) and
anti-baryons. This implies that not only mass plays an important
factor in the particle yield at a given \pT, the particle species also
significantly affects the outcome in p+p collisions. This
characteristic baryon versus meson grouping in p+p data has been seen
previously in $m_T$ scaling analyses of the same data~\cite{ppSplit},
and our results confirm this observation that baryon number plays an
important role in hadron production in p+p collisions.

\section{Discussions and Outlooks}

Modifying the Blast-Wave model to utilize Tsallis statistics instead
of the conventional Boltzmann-Gibbs statistics has allowed high
quality fits (see $\chi^2$/DoF in Table~\ref{table}) over a broader
transverse momentum range and has altered the conclusions which can be
drawn from the fits. The extended $p_T$ range is enabled by the
Tsallis statistics' capacity to evolve from an $m_T$ exponential
source into a power law through increasing $q$ values, though the
physical interpretation of this statistical model in the context of
high energy nuclear collisions remains to be fully understood.

For central Au+Au collisions at \sNN= 200 GeV, the TBW fits find a
value of $q$ approaching unity, which implies results very similar to
those from BGBW with a large radial flow velocity~
\cite{starPiKPLowPtPRL}.  The progression of the TBW fit parameters
with centrality is generally smooth but is strikingly different from
BGBW fits, culminating in a notably non-equilibrium description of \pp
and peripheral Au+Au collisions where a preference is found for zero
collective flow. Qualitatively, increasing centrality produces a
strong increase in flow velocity, a mild increase in temperature, and
a dramatic decrease in ($q-1$).  This is consistent with a picture of
increased thermalization with centrality but disfavors complete
thermal equilibrium in all systems, a requirement for applicability of
the BGBW model. The BGBW model appears to translate the
non-equilibrium features, indicated by our model, into non-zero
collective flow velocity and higher temperatures in \pp and peripheral
Au+Au collisions~\cite{Retiere_Lisa_PRC70,starPiKPLowPtPRL}

Hydrodynamics with space-time evolution from an initial
condition~\cite{hydroReview} is so far the most realistic simulation
of what happens in relativistic heavy ion collisions. However, even
with implementation of initial conditions and an interface to hadronic
cascade models at late stage of the evolution, it is hard to obtain an
intuitive picture, in contrast to that offered by an analytical
parametrization in the Blast-Wave model. It seems unlikely that
hydrodynamics is applicable at all for \pp and very peripheral A+A
collisions at RHIC. Being able to provide a systematic comparison
between p+p and central A+A collisions in one model framework is still
valuable and in some cases may be necessary.

We have additionally observed that \pp spectra continue to be well
described by our TBW curves from 3 to 10 GeV/$c$, well beyond the
range used for the fits. This implies that the origin of the power law
behavior may be the same from low \pT to high $p_T$, whether their
underlying connection is fragmentation, parton evolution, or parton
interaction cross-section. However, the high \pT behavior in central
Au+Au collisions is notably different: the experimental spectra have a
reduced power-law tail relative to the binary scaled \pp spectra, but
still significantly above the extension of the fit from $p_T<3$
GeV/$c$.  In the near future, we plan to implement a corona-like
radius-dependent ($q-1$) which could accommodate larger power-law
tails in central Au +Au collisions and allow the fit range to be
extended to higher \pT. This could provide additional information on
jet quenching as characterized by surface emission.

It has been argued~\cite{Wilk:1999dr,Wilk:2008ue} that the dependence
of $T$ and $\beta$ on ($q-1$) is related to bulk viscosity. If this
viscosity is very large at phase transition, a systematic study of TBW
model fits from AGS to SPS and RHIC energies may help locate the
critical point. This, however, requires measurements of many
identified particles to at least intermediate \pT (3 GeV/$c$) at all
energies in a systematic fashion. These data can be provided by a RHIC
beam energy scan. At LHC, one expects even larger ($q-1$) value in \pp
collisions than that at RHIC due to increased relative contributions
of hard and semi-hard processes; the conventional BGBW is not expected
to be very meaningful for \pp and peripheral Pb+Pb collisions.  If
($q-1$) is not larger for LHC \pp collisions and a non-zero flow
velocities is observed, the question of thermalization or even QGP in
such collisions might be raised. The results of TBW fits at LHC will
certainly be informative.

\section{Conclusions}

In summary, we have implemented the Tsallis statistics in a Blast-Wave
model and applied it to sets of identified particle spectra versus
transverse momenta at mid-rapidity at RHIC. This new TBW function fits
the RHIC data quite well for $p_T<$3 GeV/$c$. We observe that the
collective flow velocity starts from zero in \pp and peripheral Au+Au
collisions and rises to 0.470$\pm $0.009 $c$ in central Au+Au
collisions. The parameter ($q-1$), which characterizes the degree of
non-equilibrium in a system, changes from $0.100\pm0.003$ to
$0.015\pm0.005$ systematically from \pp to central Au+Au collisions,
indicating an evolution from a highly non-equilibrated system in
\pp collisions toward an almost thermalized system in central
Au+Au. TBW fits using all species from multiple centralities
demonstrate a quadratic dependence of the temperature and collective
flow velocity on ($q-1$), diverging into two different fits for mesons
and baryons necessary to best describe \pp spectra.

\section{Acknowledgments}

The authors would like to thank Drs. Aihong Tang, Bedanga Mohanty,
James Dunlop, Paul Sorensen, Hank Crawford and Mike Lisa for valuable
discussions.  We thank the STAR Collaboration and the RCF at BNL for
their support. This work was supported in part by the Offices of NP
and HEP within the U.S. DOE Office of Science under the contracts of
DE-FG02-88ER40412 and DE-AC02-98CH10886; Authors Yichun Xu and Zebo
Tang are supported in part by National Natural Science Foundation of
China under Grant No. 10610286 (10610285), 10475071, 10575101 and
10805046 and Knowledge Innovation Project of Chinese Academy of
Sciences under Grant No. KJCX2-YW-A14. Lijuan Ruan thanks the Battelle
Memorial Institute and Stony Brook University for the support in the
form of the Gertrude and Maurice Goldhaber Distinguished
Fellowship. Zhangbu Xu is supported in part by the PECASE Award.

%\appendix
%\section{}

%\begin{acknowledgments}
%\end{acknowledgments}


\bibliography{TBW}

\end{document}